\documentclass[prd,longbibliography,superscriptaddress,onecolumn,showpacs,showkeys,floatfix,a4paper,eqsecnum]{revtex4-1}

\usepackage{textcomp}
\usepackage{eurosym}
\usepackage{amsfonts}
\usepackage{array}
\usepackage{amsthm}
\usepackage{bm}
\usepackage{palatino}
\usepackage[colorinlistoftodos]{todonotes}
\usepackage{mathpazo}
\usepackage{supertabular}
\usepackage{subfig}
\usepackage{amssymb}
\usepackage{eurosym}
\usepackage{amsmath}
\usepackage{epsfig}
\usepackage{graphics}
\usepackage{color}
\usepackage{graphicx}
\usepackage[colorlinks=true,
            linkcolor=blue,
            urlcolor=blue,
            citecolor=green]{hyperref}
\def\be{\begin{equation}}
\def\ee{\end{equation}}
\def\beq{\begin{eqnarray}}
\def\eeq{\end{eqnarray}}

\def\bes{\begin{eqnarray}}
\def\ees{\end{eqnarray}}

\begin{document}

\title{Deflection of light by rotating regular black holes using the Gauss-Bonnet theorem }
\author{Kimet Jusufi}
\email{kimet.jusufi@unite.edu.mk}
\affiliation{Physics Department, State University of Tetovo, Ilinden Street nn, 1200,
Tetovo, MACEDONIA.}
\affiliation{Institute of Physics, Faculty of Natural Sciences and Mathematics, Ss. Cyril and Methodius University, Arhimedova 3, 1000 Skopje, MACEDONIA.}

\author{Ali \"{O}vg\"{u}n}
\email{ali.ovgun@pucv.cl}
\homepage{http://www.aovgun.com}
\affiliation{Instituto de F\'{\i}sica, Pontificia Universidad Cat\'olica de
Valpara\'{\i}so, Casilla 4950, Valpara\'{\i}so, CHILE.}

\affiliation{Physics Department, Arts and Sciences Faculty, Eastern Mediterranean University, Famagusta, North Cyprus via Mersin 10, TURKEY.}

\affiliation{Physics Department, California State University Fresno, Fresno, CA 93740, USA.} 

\affiliation{Stanford Institute for Theoretical Physics, Stanford University, Stanford,
CA 94305-4060, USA.} 

\author{Joel Saavedra}
\email{joel.saavedra@ucv.cl} 

\affiliation{Instituto de F\'{\i}sica, Pontificia Universidad Cat\'olica de
Valpara\'{\i}so, Casilla 4950, Valpara\'{\i}so, CHILE.}

\author{Yerko V\'asquez}
\email{yvasquez@userena.cl}
\affiliation{Departamento de F\'isica y Astronom\'ia, Facultad de Ciencias, Universidad de La Serena,\\
Avenida Cisternas 1200, La Serena, CHILE.}

\author{P. A. Gonz\'{a}lez}
\email{pablo.gonzalez@udp.cl} \affiliation{Facultad de
Ingenier\'{i}a y Ciencias, Universidad Diego Portales, Avenida Ej\'{e}rcito
Libertador 441, Casilla 298-V, Santiago, CHILE.}

\date{\today }

\begin{abstract}
In this paper, we study the weak gravitational lensing in the spacetime of rotating regular black hole geometries such as Ayon-Beato-Garc\'ia (ABG), Bardeen, and Hayward black holes. We calculate the deflection angle of light using the Gauss-Bonnet theorem (GBT) and show that the deflection of light can be viewed as a partially topological effect in which the deflection angle can be calculated by considering  a domain outside of the light ray applied to the black hole optical geometries. Then, we demonstrate also the deflection angle via 
the geodesics formalism 
for these black holes to verify our results and explore the differences with the Kerr solution. These black holes have in addition to the total mass and rotation parameter, different parameters as electric charge, magnetic charge, and deviation parameter. Newsworthy, we find that the deflection of light has 
correction terms coming from these parameters which generalizes the Kerr deflection angle. 
\end{abstract}

\maketitle
\tableofcontents

\section{Introduction}
Since Einstein discovered general theory of relativity in 1915 \cite{Einstein}, Einstein's theory has been subjected to numerous experimental tests. It turns out that, experimental results are quite well in agreement with theoretical predictions of this theory, starting from astrophysics observations, but also a number of other precise confirmed experiments \cite{deSitter,ligo,ligo2,Joha}. Some of the predictions are exciting: gravitational waves which were recently detected by LIGO \cite{ligo,ligo2}, gravitational lensing and bending of light, black holes, wormholes and others.  The gravitational lensing has been studied previously in the literature using different types of spacetimes with strong lensing or weak lensing. 

In this paper, we focus on the weak gravitational lensing using the Gauss-Bonnet theorem (GBT) also known as the Gibbons-Werner method (GWM). Since the black holes can not be observed directly, one way to ensure their existence is to study the geodesic equations of light rays in the curved spacetime geometry due to the presence of black holes. In this way, one can extract valuable information from black holes and detect their features. Weak gravitational lensing is an interesting method, however in most of the cases the strong lensing regime is needed; strong lensing provides more information from experimental point of view to detect other exotic objects or ultra-compact objects such as boson stars \cite{bos}. In the near future, scientists expect to detect the horizon of a black hole using the Event Horizon Telescope (EHT) \cite{event}; so, that this topic has acquired a great interest, and many authors focus on it to obtain correct results \cite{sd1,sd2,sd3,sd4,sd5,sd6,sd7,sd8,sd9,sd10,sd11,ahmed1,ahmed2,ahmed3,sd12,sd13,asada2,harada1,tsukamoto,strong1,Huang:2013jqa,Keeton:2005jd,sumanta,r1,r2,Das,Cve,Gibbons2015,Gonza1,Gonza2,Gonza3,Gonza4}.

Recently, Gibbons and Werner \cite{gibbons1} have changed the standard viewpoint related to the way we usually calculate the deflection angle. They have showed that one can calculate the  deflection angle in a very elegant way, namely they have used the GBT in the context of the optical geometry. The physical significance relies in the fact that one can view the bending of light ray as a global effect which is different from the standard viewpoint where the bending of light is usually associated within a region with a radius compared to the impact parameter. In this method, one shall only focus on a non-singular domain outside of the light ray. For asymptotically flat spacetimes the deflection angle can be calculated by the following equation:
\begin{equation}\notag
\hat{\alpha}=-\int \int_{S_\infty} \mathcal{K}  \mathrm{d}\sigma,
\end{equation} 
where $\mathcal{K}$ is the Gaussian optical curvature, and $\mathrm{d}\sigma$ is the surface element of the optical geometry. Note that the above expression for the deflection angle holds in the case of asymptotically flat spacetimes, whereas in the case of non asymptotically flat metric only a finite distance corrections can be studied.  Very recently, Werner has been able to extend this method to cover Kerr black holes using the Finsler-Randers type metric. More specifically, he has applied the Naz{\i}m's method to construct a Riemannian manifold osculating the Randers manifold \cite{werner}. In addition, this method has been extended to the wormhole geometries and non-asymptotically flat spacetimes with topological defects \cite{kimetwor}. This method has been used in a number of papers \cite{kimet1,kimet2,kimet3,kimet4,kimet5,aovgun1,kimet6,kimet7,kimet8,kimet9,Goulart}, among others we note that the GBT has been used in the interesting papers by Ishihara et al. \cite{Ishi1,Ishi2,ishi3}, in which the deflection angle for finite distances for a static (including the presence of the cosmological constant), and stationary metrics, is studied in a rather different setup.  

Classical singularities in general relativity break down the laws of physics. Singularities appear from the cosmological Big Bang theory to black holes where they are hidden behind the event horizon of a black hole. It is widely believed that quantum mechanics forbids the physics-ending singularities, but until today the problem of singularities remains an open problem in physics. To overcome the problem of singularities in black holes, many physicists have tried to construct regular black holes even in the context of classical general relativity. In this line of research we point out the Ayon-Beato-Garcia black hole (ABG) \cite{AyonBeato:1998ub,AyonBeato:1999ec}, Bardeen regular black holes \cite{bardeen} and Hayward regular black holes \cite{Hayward:2005gi}, which were obtained by finding a new mass function, and we obtain the deflection angles \cite{Toshmatov:2014nya,Ghosh,Tsukamoto:2017fxq,Bambi:2013ufa,Azreg-Ainou:2014pra,Toshmatov:2017zpr}, in order to explore the difference with the Kerr black hole solution in the weak gravitational lensing. In doing so, we extend the GBT method to rotating black holes with electric and magnetic charge for the first time.

This paper is organized as follows. In Section II we start by reviewing some of the basic concepts related to Finsler geometry. By introducing the ABG-Randers optical metric and followed by the Gaussian optical curvature we calculate the deflection angle. We study also the geodesic equation to verify our results. In Section III, we evaluate the deflection angle in the spacetime of a Bardeen black hole. In Section IV, we perform the same analysis for the Hayward regular black hole. Hence, we finalize our results in Section V.

\section{Deflection angle of rotating regular ABG black hole}

In this section, first we use the rotating Ayon-Beato-Garc\'{i}a spacetime, which is a nonsingular exact black hole solution of Einstein field equations coupled to a nonlinear
electrodynamics and satisfy the weak energy condition. The metric of the ABG black hole is written in the form \cite{Toshmatov:2014nya,Ghosh}:
\begin{equation} \label{1}
ds^2=-f(r,\theta) dt^2 +\frac{\Sigma}{\Delta}dr^2-2 a \sin^2 \theta \left(1-f(r,\theta) \right) d\phi dt+\Sigma d\theta^2+ \sin^2\theta \left[\Sigma-a^2 \left(f(r,\theta)-2\right)\sin^2\theta  \right]d \phi^2~,
\end{equation}
with 
\begin{eqnarray}
f(r,\theta) &=& 1-\frac{2 m r \sqrt{\Sigma}}{\left(\Sigma +Q^2\right)^{3/2}}+\frac{Q^2 \Sigma }{\left(\Sigma +Q^2\right)^{2}}~,  \\
\Sigma &=& r^2+a^2 \cos^2\theta~,  \\
\Delta &=& \Sigma f(r,\theta)+a^2 \sin^2\theta~. 
\end{eqnarray}

The ABG black hole metric can be further simplified by setting $\theta=\pi/2$, in that case the function $f(r)$ takes the form
\begin{eqnarray}
f(r) &=& 1-\frac{2 m r^2}{\left(r^2 +Q^2\right)^{3/2}}+\frac{Q^2 r^2 }{\left(r^2+Q^2\right)^{2}}~,\\
\Sigma &=& r^2~, \\
\Delta &=& \Sigma f(r)+a^2~.
\end{eqnarray}

We wish now to recast our ABG metric into the Finsler-Randers type metric of the general form \cite{werner}
\begin{equation}
\mathcal{F}(x, v)=\sqrt{\zeta_{ij}(x)v^{i}v^{j}}+\beta_{i}(x)v^{i}~,
\end{equation}
provided $\zeta^{ij}\beta_{i}\beta_{j}<1$, in which $\zeta_{ij}$ gives the Riemannian metric to be calculated from the ABG metric, while $\beta_{i}$ represents a one-form. If we solve Eq. \eqref{1} for the null geodesic case i.e. $\mathrm{d}s^2=0$, the problem is simplified to study the deflection of light in the equatorial plane by letting $\theta=\pi/2$. In that case, we find the following ABG-Randers optical metric 
%given by
\begin{equation}\label{777}
\mathcal{F}\left(r,\phi,\frac{\mathrm{d}r}{\mathrm{d}t},\frac{\mathrm{d}\phi}{\mathrm{d}t}\right)=\sqrt{\left[\frac{a^2 (1-f(r))^2}{f(r)^2}   +\frac{\Sigma-a^2(f(r)-2)}{f}\right]\left(\frac{\mathrm{d}\phi}{\mathrm{d}t}\right)^2+\frac{\Sigma }{\Delta f(r)}\left(\frac{\mathrm{d}r}{\mathrm{d}t}\right)^2}- \frac{a(1-f(r))}{f(r)}\frac{\mathrm{d}\phi}{\mathrm{d}t}~.
\end{equation}

The physical significance of the ABG-Randers optical metric $\mathcal{F}$ relies in the remarkable feature of the Finsler geometry, namely it provides a way to actually compute the null geodesics. In other words, there is a link of finding null geodesics in our physical metric \eqref{1} with the problem of finding the null geodesics of a ABG-Randers optical metric which can be seen by recalling the Fermat's principle. Since $\mathrm{d}t=\mathcal{F}(x,\mathrm{d}x)$, Fermat's principle of least time in the context of general relativity suggests that the null geodesics can be found from the following condition
\begin{equation}
\delta\, \int_{\gamma} \mathrm{d}t=\delta\,\int\limits_{\gamma_\mathcal{F}}\mathcal{F}(x, \dot{x})\mathrm{d}t=0~.
\end{equation}

Hence, it is clear that the Rander-Finsler metric $\mathcal{F}$ naturally appears in the problem of finding null geodesics and generalizes the Fermat's principle. The Randers-Finsler metric is characterized by the Hessian 
\begin{equation} \label{9}
g_{ij}(x,v)=\frac{1}{2}\frac{\partial^{2}\mathcal{F}^{2}(x,v)}{\partial v^{i}\partial v^{j}}~,
\end{equation}
where $x\in \mathcal{M},\ v\in T_x M$. To this end we need to apply Naz{\i}m's method which provides us to construct a Riemannian manifold  $(\mathcal{M}, \bar{g})$ that osculates the ABG-Randers manifold $ (\mathcal{M}, \mathcal{F}) $. For this purpose, we need to choose a vector field $\bar{v}$ tangent to the geodesic $\gamma_{\mathcal{F}}$, such that $\bar{v}(\gamma_{\mathcal{F}})=\dot{x}$. In that case the Hessian reads 
\begin{equation}\label{10}
\bar{g}_{ij}(x)=g_{ij}(x,\bar{v}(x))~.
\end{equation}

It is obvious that, the choice of the vector field is not unique and affect the optical metric components but, the crucial result which should be noted that a geodesic of the Randers manifold $\gamma_{\mathcal{F}}$ is also a geodesic $\gamma_{\bar{g}}$ of $(\mathcal{M},\bar{g})$ (see \cite{werner} for details):
\begin{equation}
\ddot{x}^i+\Gamma^{i}_{jk}(x,\dot{x})\dot{x}^{j}\dot{x}^{k}=\ddot{x}^i+{\bar{\Gamma}}^{i}_{jk}(x)\dot{x}^{j}\dot{x}^{k}=0~,
\end{equation}
or $\gamma_{F}=\gamma_{\bar{g}}$. One can choose the non-singular region $S_{R}\subset M$ to be bounded by the light ray $\gamma_{\mathcal{F}}$ and a curve $\gamma_{R}$ in a radial distance $R$ from the coordinate origin. Furthermore, these curves can be parameterized as follows \cite{werner}
\begin{eqnarray}
\gamma_{\mathcal{F}} &:& \,\,\,x^i(t)=\eta^i(t),\,\,\,t \in [0,l]\\
\gamma_{R}&:&\,\,\, x^i(t)=\zeta^i(t), \,\,\,t \in [0,l^{\star}]~.
\end{eqnarray}

\begin{figure}[h!]
\center
\includegraphics[width=0.5\textwidth]{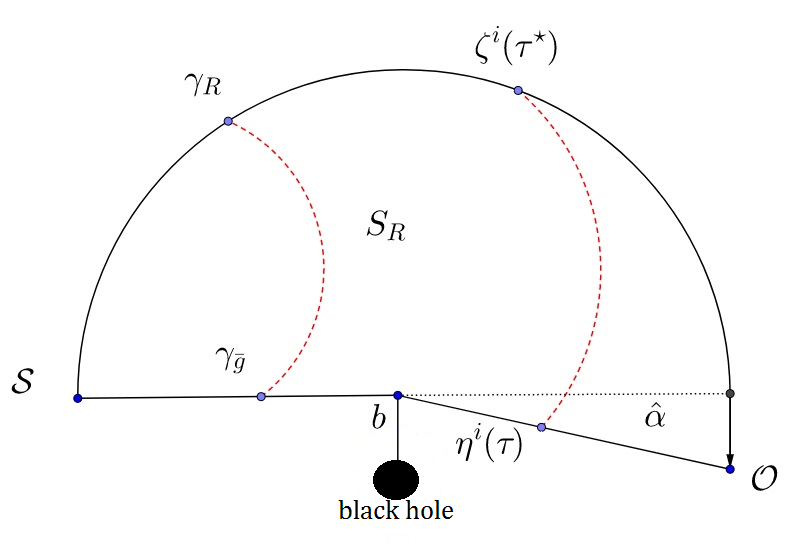}
\caption{{\protect\small \textit{The figure shows the integration domain $S_R$, namely the
equatorial plane $(r, \phi)$, in which $\hat{\alpha}$ is the total deflection angle and $b$ is the impact parameter.}}}
\label{f1}
\end{figure}

In particular one can introduce, say $\tau=t/l $, along the geodesic $\gamma_{F}$ which belongs to the interval $\in (0,1)$, and similarly $\tau^{\star}=1-t/l$ with the interval $\in (0,1)$ along the curve $\gamma_{R}$. This means that one can pair each point $\eta^i(\tau)$ on $\gamma_{\mathcal{F}}$ with ${\zeta^i}(\tau^{\star})$ on $\gamma_{R}$ if we set $\tau=\tau^\star$. In other words, one can show that there exists a family of smooth curves $x^i(\sigma,\tau)$, such that for pair of each point that there is precisely one curve which touches the boundary curve (see, Fig. \ref{f1}).

In addition we can say that $x^{i}(\sigma,\tau)$ touches the curve $\gamma_{\mathcal{F}}$ at the boundary when $\eta^i(\tau)=x^{i}(0,\tau)$, where $\sigma$ is a new parameter. Then the following relation holds
\begin{equation}
\dot{\eta}^{i}(\tau)=\frac{\mathrm{d}\eta^i}{\mathrm{d}t}(\tau)=\frac{\mathrm{d} x^i}{\mathrm{d} \sigma}(0, \tau)~.
\end{equation}

Likewise, we can say that $x^{i}(\sigma,\tau)$ touches the curve $\gamma_{R}$ when $\zeta^i(\tau)=x^{i}(1,\tau)$. This suggests that
\begin{equation} \label{15}
\dot{\zeta}^{i}(\tau)=\frac{\mathrm{d}\zeta^i}{\mathrm{d}t}(\tau)=\frac{\mathrm{d} x^i}{\mathrm{d} \sigma}(1, \tau)~.
\end{equation}

In general, one can construct a smooth and nonzero tangent vector field 
\begin{equation} \label{16}
\bar{v}^i(x(\sigma,\tau))=\frac{\mathrm{d}x^i}{\mathrm{d}\sigma} (\sigma,\tau)~,
\end{equation}
with a family of smooth curves which satisfy the following relation \cite{werner}
\begin{eqnarray}
x^{i}(\sigma,\tau)&=&\eta^i(\tau)+\dot{\eta}^{i}(\tau)\sigma+\mathcal{A}(\tau)\sigma^2+\mathcal{B}(\tau)\sigma^3 + y^i(\sigma,\tau)(1-\sigma)^2 \sigma^2~,
\end{eqnarray}
with
\begin{eqnarray}\notag
\mathcal{A}(\tau)&=& 3\zeta^{i}(\tau)-3\eta^{i}(\tau) -\dot{\zeta}^{i}(\tau)-2 \dot{\eta}^{i}(\tau)\eta^{i}(\tau),\\\notag
\mathcal{B}(\tau)&=& 2 \eta^{i}(\tau)-2 \zeta^{i}(\tau)+\dot{\zeta}^{i}(\tau)+\dot{\eta}^{i}(\tau)~.
\end{eqnarray}

That being said, and keeping in mind that our metric is asymptotically flat, we can choose an equation for the light rays as follows
\begin{equation}\label{17}
r(\phi)=\frac{b}{\sin\phi}~,
\end{equation}
with $b$ being the impact parameter. From the light ray equation one can deduce the following components for the vector field  
\begin{equation} \label{18}
\bar{v}^{r}=\frac{\mathrm{d}r}{\mathrm{d}t}=-\cos\phi,\,\,\,\bar{v}^{\phi}=\frac{\mathrm{d}\phi}{\mathrm{d}t}=\frac{\sin^{2}\phi}{b}~.
\end{equation}

It is worth noting that the choose of the vector field is dictated by the light ray equation $r_\gamma$. Note that our particular equation of the light ray represents a straight line approximation, this  will be important in the final result for the deflection angle. In other words, due to the straight line approximation we expect our deflection angle to be correct in leading order terms. 

\subsection{Gaussian optical curvature}

We shall now continue to compute the metric components. To this end, we need to combine Eqs. \eqref{9}, \eqref{18} yielding the following non-zero components
\begin{eqnarray}\notag
\bar{g}_{rr}&=& - \frac{2 \left[-2 (m+\frac{r}{4}) \left(\sin^4\phi r^2 +b^2 \cos^2 \phi \right)^{3/2}+a m r^2 \sin^6 \phi    \right]}{r \left(\sin^4\phi r^2 +b^2 \cos^2 \phi \right)^{3/2}} \\
&-& \frac{\left[ 18 (m+\frac{r}{9})\left(\sin^4\phi r^2 +b^2 \cos^2 \phi \right)^{5/2} -7 \left(13(m+\frac{r}{13})r^2 \sin^4\phi+b^2 \cos^2\phi (m+\frac{r}{7})    \right)r^2 \sin^6\phi \right]Q^2}{r^3 \left(\sin^4\phi r^2 +b^2 \cos^2 \phi \right)^{5/2}}~, \\
\bar{g}_{r \varphi}&=& \frac{2 a \cos^3 \phi m}{r \left( \frac{\sin^4\phi r^2 +b^2 \cos^2 \phi}{b^2}  \right)^{3/2}}-\frac{b^3 a Q^2 \cos^3\phi \left(17 \sin^4 \phi m r^{19}+\sin^4\phi r^{20}+11 r^{17} \cos^2 \phi b^2 m+\cos^2 \phi b^2 r^{18}   \right) }{ r^{20} \left(\sin^4\phi r^2 +b^2 \cos^2 \phi \right)^{5/2} }~,\\\notag
\bar{g}_{\varphi\varphi}&=& \frac{2r \left[-2amr^2 \sin^6\phi-3a b^2 m \cos^2\phi \sin^2\phi +(m+\frac{r}{2}) \left(\sin^4\phi r^2 +b^2 \cos^2 \phi \right)^{3/2}  \right]}{\left(\sin^4\phi r^2 +b^2 \cos^2 \phi \right)^{3/2}}\\
&-& \frac{Q^2 \left[7(m+\frac{r}{7})\left(\sin^4\phi r^2 +b^2 \cos^2 \phi \right)^{5/2}-3a \sin^2\phi \,\Xi(r,a,m)  \right]}{r\left(\sin^4\phi r^2 +b^2 \cos^2 \phi \right)^{5/2} }~,
\end{eqnarray}
where 
\begin{equation}
\Xi=6 \left(m+\frac{r}{9} \right)r^4 \sin^8 \phi +15 \cos^2 \phi \left(m+\frac{r}{9} \right)b^2 r^2 \sin^4 \phi+7 b^4 \cos^4 \phi \left(m+\frac{r}{7} \right)~.
\end{equation}

The determinant is given as
\begin{eqnarray}\notag
\det \bar{g}&=&-\frac{6 r^4 ma \sin^6\phi +6 b^2 a mr^2 \cos^2 \phi \sin^2 \phi+33 \left(   -\frac{2}{11}mr^2 -\frac{1}{33}r^3 \right)\left(\sin^4\phi r^2 +b^2 \cos^2 \phi \right)^{3/2}}{r \left(\sin^4\phi r^2 +b^2 \cos^2 \phi \right)^{3/2}}\\
&-& \frac{Q^2 \left[-51 r^2 (m+\frac{r}{11}) a \sin^6 \phi -45 b^2 a \cos^2 \phi (m+\frac{r}{15}) \sin^2 \phi +33 (m+\frac{r}{11})  \left(\sin^4\phi r^2 +b^2 \cos^2 \phi \right)^{3/2}  \right]}{r  \left(\sin^4\phi r^2 +b^2 \cos^2 \phi \right)^{3/2}}~.
\end{eqnarray}

The Gaussian optical curvature then can be found by noticing the relation $\bar{R}_{r\phi r\phi}=\mathcal{K}\,\det \bar{g}$. In other words, we can compute $\mathcal{K}$ as follows
\begin{equation}
\mathcal{K}=\frac{1}{\sqrt{\det \bar{g}}}\left[\frac{\partial}{\partial \phi}\left(\frac{\sqrt{\det \bar{g}}}{\bar{g}_{rr}}\,\bar{\Gamma}^{\phi}_{rr}\right)-\frac{\partial}{\partial r}\left(\frac{\sqrt{\det \bar{g}}}{\bar{g}_{rr}}\,\bar{\Gamma}^{\phi}_{r\phi}\right)\right]~.
\end{equation}

Our computation reveals the following result 
\begin{equation}\label{23}
\mathcal{K}=-2\,{\frac {m}{{r}^{3}}}+3\,{\frac {{Q}^{2}}{{r}^{4}}}+12\,{\frac {{Q}
^{2}m}{{r}^{5}}}+\frac{15 a \mathcal{G}(r,\phi)}{r^5}~.
\end{equation}

Note that the first term corresponds to the Schwarzschild black hole, while the second and third term give the charge contribution, finally the last term is a consequence of the rotation. Note that the function $\mathcal{G}(r,\varphi)$ is rather complicated expression which is found to be 
\begin{widetext}
\begin{eqnarray}\notag
\mathcal{G}(r,\phi)&=& - \frac{\sin^2\phi}{\left(\sin^4\phi r^2 +b^2 \cos^2 \phi \right)^{9/2}} \Big[ \left( {\frac {11\,{Q}^{2}m{r}^{8}}{10}}+\frac{1}{2}{Q}^{2}{r}^{9}-\frac{2}{5}m{r
}^{10} \right)  \left( \sin \left( \phi \right)  \right) ^{16}\\\notag
&-& \frac{1}{2}{r}^{6}{b}^{2} \left( {Q}^{2}m+\frac{1}{5}\,{Q}^{2}r-\frac{2}{5}m{r}^{2}
 \right)  \left( \sin \left( \phi \right)  \right) ^{14}\\\notag
 &+& {r}^{6}{b}^{2} \left( \cos \left( \phi \right)  \right) ^{2} \left( {Q
}^{2}m+{\frac {11\,{Q}^{2}r}{10}}+\frac{2}{5}m{r}^{2} \right)  \left( \sin
 \left( \phi \right)  \right) ^{12}\\\notag
 &+& {\frac {42\,{b}^{3}{r}^{5} \left( \cos \left( \phi \right)  \right) ^{
2} \left( \sin \left( \phi \right)  \right) ^{11}}{5} \left( {Q}^{2}m+
{\frac {5\,{Q}^{2}r}{42}}-{\frac {4\,m{r}^{2}}{21}} \right) }\\\notag
&+& \frac{9{r}^{4}{b}^{2}}{2} \left( \cos \left( \phi \right)  \right) ^{2}
 \left( \sin \left( \phi \right)  \right) ^{10} \\\notag
 &\times& \left(  \left( -{\frac {10\,{Q}^{2}m{r}^{2}}{9}}-\frac{2}{9}{Q}^{2}{r}^{3}+
\frac{4}{9}{r}^{4}m \right)  \left( \cos \left( \phi \right)  \right) ^{2}+{
b}^{2} \left( {Q}^{2}m+\frac{1}{15}{Q}^{2}r-\frac{2}{15}m{r}^{2} \right)  \right)\\\notag
 &+& {\frac {84\,{r}^{5}{b}^{3} \left( \cos \left( \phi \right)  \right) ^{
4} \left( \sin \left( \phi \right)  \right) ^{9}}{5} \left( {Q}^{2}m+{
\frac {5\,{Q}^{2}r}{42}}-{\frac {4\,m{r}^{2}}{21}} \right) }\\\notag
&+& {\frac {96\,{r}^{4}{b}^{4} \left( \cos \left( \phi \right)  \right) ^{
4} \left( \sin \left( \phi \right)  \right) ^{8}}{5} \left( {Q}^{2}m+\frac{1}{8}{Q}^{2}r-{\frac {5\,m{r}^{2}}{96}} \right) }\\\notag
&-& {\frac {21\,{r}^{3}{b}^{5} \left( \cos \left( \phi \right)  \right) ^{
4} \left( \sin \left( \phi \right)  \right) ^{7}}{5} \left( {Q}^{2}m-{
\frac {5\,{Q}^{2}r}{21}}+\frac{2}{7}m{r}^{2} \right) }\\\notag
&-& 2\,{r}^{2}{b}^{4} \left( \cos \left( \phi \right)  \right) ^{4}
 \left( -9\,{Q}^{2}m{r}^{2} \left( \cos \left( \phi \right)  \right) ^
{2}+{b}^{2} \left( {Q}^{2}m-\frac{1}{5}{Q}^{2}r+\frac{2}{5}m{r}^{2} \right) 
 \right)  \left( \sin \left( \phi \right)  \right) ^{6}\\\notag
 &-& {\frac {42\,{r}^{3}{b}^{5} \left( \cos \left( \phi \right)  \right) ^{
6} \left( \sin \left( \phi \right)  \right) ^{5}}{5} \left( {Q}^{2}m-{
\frac {5\,{Q}^{2}r}{21}}+\frac{2}{7}m{r}^{2} \right) }\\\notag
&-& {\frac {7\,{r}^{2}{b}^{6} \left( \cos \left( \phi \right)  \right) ^{6
} \left( \sin \left( \phi \right)  \right) ^{4}}{10} \left( {Q}^{2}m-{
\frac {20\,{Q}^{2}r}{7}}+{\frac {18\,m{r}^{2}}{7}} \right) }\\\notag
&+& \frac{1}{5} \left( 7\,{Q}^{2}+2\,{r}^{2} \right) r{b}^{7} \left( \cos
 \left( \phi \right)  \right) ^{6}m \left( \sin \left( \phi \right) 
 \right) ^{3}+{Q}^{2}{b}^{8} \left( \cos \left( \phi \right)  \right) ^{8} \left( m+
r/5 \right)\\\notag
 &-& 5\,{r}^{2}{b}^{6} \left( \cos \left( \phi \right)  \right) ^{8}
 \left( {Q}^{2}m-\frac{1}{5}{Q}^{2}r+\frac{2}{5}m{r}^{2} \right)  \left( \sin
 \left( \phi \right)  \right) ^{2}\\
 &+& \frac{1}{5} \left( 14\,{Q}^{2}+4\,{r}^{2} \right) r{b}^{7} \left( \cos
 \left( \phi \right)  \right) ^{8}m\sin \left( \phi \right) 
\Big]~.
\end{eqnarray}
\end{widetext}

Note that the Gaussian optical curvature depends on the black hole parameters, $a$, $m$, and $Q^2$. In the next section we are going to evaluate the deflection angle with the help of the above result.

\subsection{Deflection angle}
\textbf{Theorem:}
\textit{Let $(S_{R},\bar{g})$ be a non-singular and simply connected domain over the osculating Riemannian manifold $(\mathcal{M},\bar{g})$ bounded by circular curve $\gamma_ {R}$ and the geodesic $\gamma_{\bar{g}}$. Let $\mathcal{K}$ be the Gaussian curvature of $(\mathcal{M},\bar{g})$, and $\kappa$ the geodesic curvature of $\partial S_{R}=\gamma_{\bar{g}}\cup \gamma_ {R}$. Then, the GBT can be stated as follows} \cite{gibbons1,werner}
\begin{equation}\label{19-4}
\iint\limits_{S_{R}}\mathcal{K}\,\mathrm{d}\sigma+\oint\limits_{\partial S_{R}}\kappa\,\mathrm{d}t+\sum_{k}\alpha_{k}=2\pi\chi(S_{R})~.
\end{equation}

As we have already noted that $\mathrm{d}\sigma$ gives the surface element, $\alpha_{k}$ represents the $k^{th}$ exterior angles, $\chi(S_{R})$ is known as the Euler characteristic number. The geodesic curvature basically determines the deviation from the geodesic. By definition we have $\kappa(\gamma_{\bar{g}})=0$, because $\gamma_{\bar{g}}$ is a geodesic. Of particular importance is the geodesic curvature of $\gamma_R$ in a radial coordinate $R$ from the coordinate origin. It can be calculated via 
\begin{equation}
\kappa (\gamma_{R})=|\nabla _{\dot{\gamma}_{R}}\dot{\gamma}_{R}|~.
\end{equation}

We can choose $\gamma_{R}:=r(\varphi)=R=\text{const}$, in that case the radial part yields
\begin{equation}
\left( \nabla _{\dot{\gamma}_{R}}\dot{\gamma}_{R}\right) ^{r}=\dot{\gamma}_{R}^{\phi
}\,\left( \partial _{\phi }\dot{\gamma}_{R}^{r}\right) +\bar{\Gamma} _{\phi
\phi }^{r}\left( \dot{\gamma}_{R}^{\phi }\right) ^{2}~. \label{12}
\end{equation}

It is noted that the first term vanishes, while the second term can be calculated by the unit speed condition i.e., $\bar{g}_{\phi \phi} \dot{\gamma}_{R}^{\phi } \dot{\gamma}_{R}^{\phi }=1$. Since our optical geometry is asymptotically Euclidean we find that $\kappa(\gamma_{R}) \to R^{-1}$ as $R\to \infty $. The other point is the fact that as $R\rightarrow \infty $, the sum of jump angles ($\alpha _{\mathcal{O}}$), to the source $\mathcal{S}$, and observer $\mathcal{O}$, yields $\alpha_{\mathit{O}%
}+\alpha_{\mathit{S}}\rightarrow \pi $ \cite{gibbons1}. For constant $R$, the optical metric gives
\begin{eqnarray}\notag
\lim_{R\to \infty}\mathrm{d}t &=&\lim_{R\to \infty}\left[ \sqrt{\frac{a^2 (1-f(r))^2}{f(r)^2}   +\frac{\Sigma-a^2(f(r)-2)}{f}}- \frac{a(1-f(r))}{f(r)}\right]\mathrm{d}\phi\\
&\to & R \mathrm{d}\phi~,
\end{eqnarray}
where we have used the fact that 
\begin{equation}
\lim_{R \to \infty} f(R) \to 1~.
\end{equation}
Finally one can shows that
\begin{equation}
\lim_{R \to \infty} \kappa(\gamma_{R})\frac{\mathrm{d}t}{\mathrm{d} \phi}\to 1~.
\end{equation}

Note that by construction, the source $S$ and the observer $O$ are assumed to be in the asymptotically Euclidean region, thus the last equation clearly reveals our assumptions that our optical metric is asymptotically Euclidean. Having computed the geodesic curvature from GBT it follows
\begin{equation}
\iint\limits_{\mathcal{S}_{R}}\mathcal{K}\,\mathrm{d}S+\oint\limits_{\gamma_{R}}\kappa \,
\mathrm{d}t\overset{{R\rightarrow \infty }}{=}\iint\limits_{\mathcal{S}%
_{\infty }}\mathcal{K}\,\mathrm{d}\sigma+\int\limits_{0}^{\pi +\hat{\alpha}}\mathrm{d}\phi
=\pi~,
\end{equation}
 resulting with 
\begin{equation}
\hat{\alpha}=-\iint\limits_{\mathcal{S}_{\infty }}\mathcal{K}\mathrm{d}\sigma~.
\end{equation}

After substituting the Gaussian optical curvature \eqref{23} into the last equation we find 
\begin{equation}
\hat{\alpha} \simeq  -\int\limits_{0}^{\pi}\int\limits_{\frac{b}{\sin \phi}}^{\infty}\left[-2\,{\frac {m}{{r}^{3}}}+3\,{\frac {{Q}^{2}}{{r}^{4}}}+12\,{\frac {{Q}
^{2}m}{{r}^{5}}}+\frac{15 a \mathcal{G}(r,\phi)}{r^5}\right]\,\sqrt{\det \bar{g}}\,\mathrm{d}r\,\mathrm{d}\phi~.
\end{equation}

Solving the non-rotating part in the above integral we find
\begin{eqnarray}\notag
\mathcal{I}_{1}&=&-\int\limits_{0}^{\pi}\int\limits_{\frac{b}{\sin \varphi}}^{\infty}\left( -2\,{\frac {m}{{r}^{3}}}+3\,{\frac {{Q}^{2}}{{r}^{4}}}+12\,{\frac {{Q}
^{2}m}{{r}^{5}}}\right)\sqrt{\det \bar{g}}\mathrm{d}r\mathrm{d}\varphi\\
&=& \frac{4m}{b}-\frac{3 \pi Q^2}{4 b^2}-\frac{16 Q^2 m}{3 b^3}~.
\end{eqnarray}

The rotating part gives
\begin{equation}
\mathcal{I}_2= -\int\limits_{0}^{\pi}\int\limits_{\frac{b}{\sin \varphi}}^{\infty}\left(\frac{15 a \mathcal{G}(r,\phi)}{r^5} \right)\,\sqrt{\det \bar{g}}\,\mathrm{d}r\,\mathrm{d}\varphi=\pm \frac{4ma}{b^2} \pm \frac{98 96 ma Q^2}{15 \,b^4}~.
\end{equation}

The total deflection angle is found 
\begin{equation} \label{deflectionangle}
\hat{\alpha} = \frac{4m}{b}-\frac{3 \pi Q^2}{4 b^2} \pm \frac{4ma}{b^2} +\mathcal{O}(Q^2,a,m)~,
\end{equation}
where the signs of positive and negative stand for a retrograde and a prograde light rays.

\subsection{Geodesics}

The equations of the geodesics can be derived from the Lagrangian of a test particle \cite{chandra}. For motion in the equatorial plane, that is, $\theta=\pi/2$ and $\dot{\theta}=0$, the Lagrangian results to be:
\begin{equation}
\label{LGarcia}
  2\mathcal{L}= -f(r) \dot{t}^2 +\frac{r^2}{\Delta}\dot{r}^2-2 a \left(1-f(r) \right) \dot{\phi} \dot{t}+  \left[r^2-a^2 \left(f(r)-2\right)  \right]\dot {\phi}^2 ~,
\end{equation}
where $\dot{q}=dq/d\tau$, and $\tau$ is an affine parameter along the geodesic. Since the Lagrangian (\ref{LGarcia}) is
independent of the cyclic coordinates ($t,\phi$), then their
conjugate momenta ($\Pi_t, \Pi_{\phi}$) are conserved. Then, the equations of motion are obtained from
$ \dot{\Pi}_{q} - \frac{\partial \mathcal{L}}{\partial q} = 0$, and yield
\begin{equation}
\dot{\Pi}_{t} =0 , \quad \dot{\Pi}_{\phi}=0~,
\label{w.11a}
\end{equation}
where $\Pi_{q} = \partial \mathcal{L}/\partial \dot{q}$
are the conjugate momenta to the coordinate $q$, and are given by
\begin{equation}
\Pi_{t} = -f(r)\dot{t}-a(1-f(r))\dot{\phi}\equiv -E~, 
\quad \Pi_{r}=\frac{r^2}{\Delta}\dot{r}~\quad\textrm{and}
\quad \Pi_{\phi}
=-a(1-f(r))\dot{t}+(r^2-a^2(f(r)-2))\dot{\phi}\equiv L~,
\label{w.11c}
\end{equation}
where $E$ and $L$ are dimensionless integration constants associated to each of them. So, the Hamiltonian is given by
\begin{equation}
\mathcal{H}=\Pi_{t} \dot{t} + \Pi_{\phi}\dot{\phi}+\Pi_{r}\dot{r}
-\mathcal{L}
\end{equation}
\begin{equation}
2\mathcal{H}=-E\, \dot{t} + L\,\dot{\phi}+\frac{r^2}{\Delta}\dot{r}^2\equiv -\bar{m}^2~.
\label{w.11z}
\end{equation}

Now, by normalization, we shall consider $\bar{m}^2 = 0$ for photons. Therefore, we obtain   

\begin{eqnarray}
\label{ABG1}
&&\dot{t}= \frac{2a^2E-aL+Er^2+a(L-aE)f(r)}{a^2+r^2f(r)}~,\\
\label{ABG2}
&&\dot{\phi}= \frac{aE+f(r)(L-aE)}{a^2+r^2f(r)}~,\\
\label{ABG3}
&&\dot{r}^{2}=  \frac{a(2aE^2-2EL)+E^2r^2-f(r)(L-aE)^2}{r^2}~.
\end{eqnarray}

The distance of the closest approach $r_0$ for the metric (\ref{1}) can be obtained from $\dot{r}=0$, which yields 
\begin{equation}
\label{ipABG}
\frac{r_0}{b}=\sqrt{1-\left(\frac{a}{b}\right)^2+r_0^2\left(1-\frac{a}{b}\right)^2\left(\frac{Q^2}{(Q^2+r_0^2)^2}-\frac{2m}{(Q^2+r_0^2)^{3/2}}\right)}~,
\end{equation}
where $b=L/E$ is the impact parameter. 

\begin{figure}[h]
\begin{center}
\includegraphics[width=0.4\textwidth]{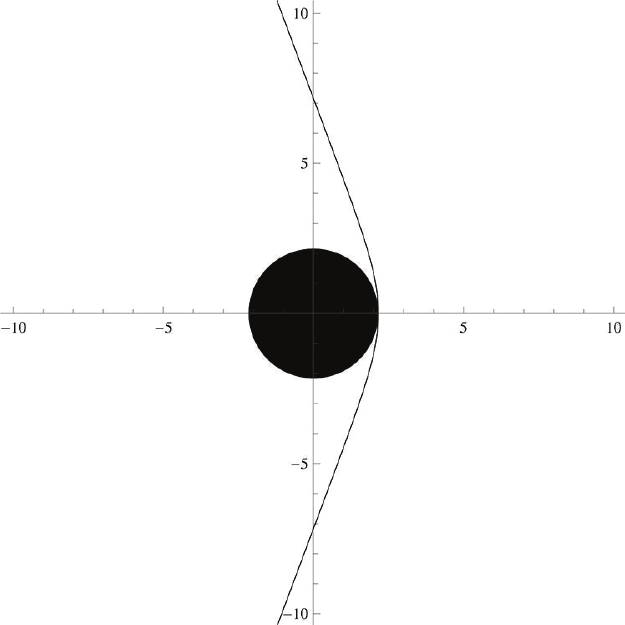}
\end{center}
\caption{The deflection of light in the background of a rotating regular ABG black hole with $E=2$, $L=5$, $m=0.3$, $Q=0.1$ and $a=0.1$. The circle corresponds to the closest approach ($r_0=2.158$) to the black hole.} \label{plotABG}
\end{figure}

Now, following \cite{Keeton:2005jd} the bending angle can be determined by the expression 
\begin{equation}
\label{angle}
\hat{\alpha}=2\int_{r_0}^{\infty}\left|\frac{d\phi}{dr}\right|dr-\pi~,
\end{equation}
which yields
\begin{equation}
\hat{\alpha}=\frac{4m}{b}-\frac{3 \pi Q^2}{4 b^2} \pm \frac{4ma}{b^2} +\mathcal{O}(Q^2,a,m)~,
\end{equation}
where we use the change of variables $u=r_0/r$; then, we substitute the impact parameter given by the Eq. (\ref{ipABG}), and we expand in Taylor series around $m$, $a$, and $Q$. Finally, we consider $r_0\approx b$. In Fig. \ref{plotABG} we plot the deflection of light in the background of a rotating regular ABG black hole by solving numerically the Eqs.(\ref{ABG2}) and (\ref{ABG3}).  

\section{Deflection angle by rotating regular Bardeen
black hole}
In this section, we study the deflection angle by rotating Bardeen regular black hole. The spacetime metric of the rotating Bardeen regular black hole reads \cite{Ghosh}:
\begin{equation}\label{metricBardeen}
ds^2=-\left(1-\frac{2\mathcal{M}_b r}{\Sigma}\right)dt^2-\frac{4 a\mathcal{M}_b r \sin^2\theta }{\Sigma }dt d\phi+\frac{\Sigma}{\Delta}dr^2+\Sigma d\theta^2+\left( r^2+a^2+\frac{2 a^2 \mathcal{M}_b r \sin^2\theta}{\Sigma}\right) d\phi^2~,
\end{equation} 
where 
\begin{eqnarray}
\Sigma  &=& r^2+a^2\cos^2\theta~,\\
\Delta  &=& r^2-2\mathcal{M}_b r+a^2 ~,\\
\mathcal{M}_b &=& m\left( \frac{r^2}{r^2+g_{\star}^2} \right)^{3/2}~,
\end{eqnarray}
in which $ g_{\star}$ is the magnetic charge due to the non-linear electromagnetic field. In this case, 
we find the following expression for the optical metric
\begin{equation}\label{7}
\mathcal{F}\left(r,\phi,\frac{\mathrm{d}r}{\mathrm{d}t},\frac{\mathrm{d}\phi}{\mathrm{d}t}\right)=\sqrt{\frac{r^4}{\Delta(\Delta-a^2)}(\frac{\mathrm{d}r}{\mathrm{d}t})^2+\frac{r^4 \Delta}{\Delta-a^2}(\frac{\mathrm{d}\phi}{\mathrm{d}t})^2}- \frac{2 m a r}{\Delta-a^2}\left( \frac{r^2}{r^2+g_{\star}^2} \right)^{3/2}\frac{\mathrm{d}\phi}{\mathrm{d}t}~,
\end{equation}

\begin{eqnarray}\notag
\bar{g}_{rr}&=& - \frac{2 \left[-2 (m+\frac{r}{4}) \left(\sin^4\phi r^2 +b^2 \cos^2 \phi \right)^{3/2}+a m r^2 \sin^6 \phi    \right]}{r \left(\sin^4\phi r^2 +b^2 \cos^2 \phi \right)^{3/2}}-\frac{g_{\star}^2 m \left[3 a r^2 \sin^6\phi +6 \left(\sin^4\phi r^2 +b^2 \cos^2 \phi \right)^{3/2}\right] }{r^3 \left(\sin^4\phi r^2 +b^2 \cos^2 \phi \right)^{3/2}}~,\\
%&-& \\\notag
 \bar{g}_{r\phi}&=&  \frac{2 a b^3 \cos^3 \phi m}{r \left( \sin^4\phi r^2 +b^2 \cos^2 \phi  \right)^{3/2}} - \frac{3 a b^3 \cos^3 \phi m g_{\star}^2}{r^3 \left( \sin^4\phi r^2 +b^2 \cos^2 \phi  \right)^{3/2}}~, \\\notag
 \bar{g}_{\phi \phi}&=& \frac{2r \left[-2amr^2 \sin^6\phi-3a b^2 m \cos^\phi \sin^2\phi +(m+\frac{r}{2}) \left(\sin^4\phi r^2 +b^2 \cos^2 \phi \right)^{3/2}  \right]}{\left(\sin^4\phi r^2 +b^2 \cos^2 \phi \right)^{3/2}} \\
&-& \frac{3 g_{\star}^2 m \left[-2 a r^2 \sin^6\phi-3a b^2  \cos^2\phi \sin^2\phi+\left(\sin^4\phi r^2 +b^2 \cos^2 \phi \right)^{3/2}    \right]}{r \left(\sin^4\phi r^2 +b^2 \cos^2 \phi \right)^{3/2}}~.
\end{eqnarray}

With the determinant
\begin{eqnarray}\notag
\det \bar{g}&=&r^2- \frac{6 r m \left[a \sin^2\phi-\sqrt{ \cos^4\phi r^2+(b^2-2r^2)\cos^2 \phi +r^2}    \right]}{\sqrt{ \cos^4\phi r^2+(b^2-2r^2)\cos^2 \phi +r^2}}\\
&+&\frac{9 g_{\star}^2 m \left[a \sin^2\phi-\sqrt{ \cos^4\phi r^2+(b^2-2r^2)\cos^2 \phi +r^2}    \right]}{r\sqrt{ \cos^4\phi r^2+(b^2-2r^2)\cos^2 \phi +r^2}}~.
\end{eqnarray}

Then the Gaussian optical curvature is
\begin{equation}
\mathcal{K}=-\frac{2m}{r^3}+\frac{18 m g_{\star}^2}{r^5}+\frac{27 a m  \mathcal{S}(r,\phi)}{r^{5}}~,
\end{equation}
with
\begin{eqnarray}\notag
\mathcal{S}(r,\phi)&=&- \frac{\sin^2\phi}{\left(\sin^4\phi r^2 +b^2 \cos^2 \phi \right)^{7/2}}\Big[ \left( {\frac {14\,g_{\star}^{2}{r}^{6}}{9}}-2/9\,{r}^{8} \right)  \left( 
\sin \left( \phi \right)  \right) ^{12}\\\notag
&-& 1/6\,{b}^{2}{r}^{4} \left( g_{\star}^{2}-2/3\,{r}^{2} \right)  \left( \sin
 \left( \phi \right)  \right) ^{10}+ {\frac {31\,{b}^{2}{r}^{4} \left( \cos \left( \phi \right)  \right) ^{
2} \left( \sin \left( \phi \right)  \right) ^{8}}{9} \left( g_{\star}^{2}+{
\frac {4\,{r}^{2}}{31}} \right) }\\\notag
&+&  \left( 2\,{b}^{3}g_{\star}^{2}{r}^{3}-{\frac {8\,{b}^{3}{r}^{5}}{9}}
 \right)  \left( \cos \left( \phi \right)  \right) ^{2} \left( \sin
 \left( \phi \right)  \right) ^{7}\\\notag
 &+& 1/3\, \left( -5\,{r}^{2} \left( \cos \left( \phi \right)  \right) ^{2}
+2\,{b}^{2} \right) {b}^{2}{r}^{2} \left( \cos \left( \phi \right) 
 \right) ^{2} \left( g_{\star}^{2}-2/3\,{r}^{2} \right)  \left( \sin \left( 
\phi \right)  \right) ^{6}\\\notag
&+&  \left( 4\,{b}^{3}g_{\star}^{2}{r}^{3}-{\frac {16\,{b}^{3}{r}^{5}}{9}}
 \right)  \left( \cos \left( \phi \right)  \right) ^{4} \left( \sin
 \left( \phi \right)  \right) ^{5}+1/18\, \left( 97\,g_{\star}^{2}-18\,{r}^{2} \right) {b}^{4}{r}^{2} \left( 
\cos \left( \phi \right)  \right) ^{4} \left( \sin \left( \phi
 \right)  \right) ^{4}\\\notag
 &+& 1/3\,{b}^{5}r \left( \cos \left( \phi \right)  \right) ^{4} \left( g_{\star}
^{2}+2/3\,{r}^{2} \right)  \left( \sin \left( \phi \right)  \right) ^{
3}+\left( \cos \left( \phi \right)  \right) ^{6}{b}^{6}{g}^{2}\\
&+& 5/3\,{b}^{4}{r}^{2} \left( \cos \left( \phi \right)  \right) ^{6}
 \left( g_{\star}^{2}-2/3\,{r}^{2} \right)  \left( \sin \left( \phi \right) 
 \right) ^{2}+2/3\,{b}^{5}r \left( \cos \left( \phi \right)  \right) ^{6} \left( g_{\star}
^{2}+2/3\,{r}^{2} \right) \sin \left( \phi \right)
  \Big]~.
\end{eqnarray}

Substituting these relations from GBT it follows the integral
\begin{equation}
\hat{\alpha} \simeq  -\int\limits_{0}^{\pi}\int\limits_{\frac{b}{\sin \phi}}^{\infty}\left( -\frac{2m}{r^3}+\frac{18 m g_{\star}^2}{r^5} +\frac{27 a m \mathcal{S}(r,\phi)}{r^{5}}\right)\,\sqrt{\det \bar{g}}\,\mathrm{d}r\,\mathrm{d}\phi~.
\end{equation}

We can split this integral in two parts. The non-rotating contribution yields
\begin{eqnarray}\notag
\mathcal{I}_{1}&=&-\int\limits_{0}^{\pi}\int\limits_{\frac{b}{\sin \varphi}}^{\infty}\left( -\frac{2m}{r^3}+\frac{18 m g_{\star}^2}{r^5}\right)\sqrt{\det \bar{g}}\mathrm{d}r\mathrm{d}\phi\\
&=& \frac{4m}{b}-\frac{8 g_{\star}^2 m}{b^3}~.
\end{eqnarray}

For the second integral we find
\begin{equation}
\mathcal{I}_2= -\int\limits_{0}^{\pi}\int\limits_{\frac{b}{\sin \phi}}^{\infty}\left( \frac{\mathcal{S}(r,\phi)}{4r^{11}}\right)\,\sqrt{\det \bar{g}}\,\mathrm{d}r\,\mathrm{d}\phi=\pm \frac{4ma}{b^2}\mp \frac{24 ma g_{\star}^2}{b^4}~.
\end{equation}

Finally the total deflection angle of rotating Bardeen regular black hole is
\begin{equation} 
\hat{\alpha} = \frac{4m}{b}-\frac{8 g_{\star}^2 m}{b^3}\pm \frac{4ma}{b^2} +\mathcal{O}(m,a,g_{\star}^2).
\end{equation}

\subsection{Geodesics equations}

In this case, the Lagrangian associated with the motion of particles in the equatorial plane ($\theta=\pi/2$ and $\dot{\theta}=0$) results to be:
\begin{equation}
\label{tl4}
  2\mathcal{L}=-\left(1-\frac{2\mathcal{M}_b}{r}\right)\dot{t}^2-\frac{4 a\mathcal{M}_b}{r }\dot{t} \dot{\phi}+\frac{r^2}{\Delta}\dot{r}^2+\left( r^2+a^2+\frac{2 a^2 \mathcal{M}_b}{r}\right) \dot{\phi}^2~,
\end{equation}
where $\dot{q}=dq/d\tau$, and $\tau$ is an affine parameter along the geodesic. 
%that
%we choose as the proper time. 
Since the Lagrangian (\ref{tl4}) is
independent of the cyclic coordinates ($t,\phi$), then their
conjugate momenta ($\Pi_t, \Pi_{\phi}$) are conserved. Then, the equations of motion can be obtained from
$ \dot{\Pi}_{q} - \frac{\partial \mathcal{L}}{\partial q} = 0$, and we obtain
\begin{equation}
\dot{\Pi}_{t} =0~, \dot{\Pi}_{\phi}=0~,
\label{w.11aa}
\end{equation}
where $\Pi_{q} = \partial \mathcal{L}/\partial \dot{q}$
are the conjugate momenta to the coordinate $q$, and are given by
\begin{equation}
\Pi_{t} = -\left( 1-\frac{2 \mathcal{M}_b}{r} \right) \dot{t} -\frac{2a \mathcal{M}_b}{r} \dot{\phi} \equiv -E~, \quad \Pi_{r}= \frac{r^2}{\Delta} \dot{r}~\textrm{and}\quad \Pi_{\phi}
=-\frac{2 a \mathcal{M}_b}{r}\dot{t} + \left( r^2+a^2 +\frac{2 a^2 \mathcal{M}_b}{r} \right) \dot{\phi} \equiv L~,
\label{w.11cc}
\end{equation}

where $E$ and $L$ are dimensionless integration constants associated to each of them.
Therefore, the Hamiltonian is given by
\begin{equation}
\mathcal{H}=\Pi_{t} \dot{t} + \Pi_{\phi}\dot{\phi}+\Pi_{r}\dot{r}
-\mathcal{L}
\end{equation}
\begin{equation}
2\mathcal{H}=-E\, \dot{t} + L\,\dot{\phi}+\frac{r^2}{\Delta} \dot{r} ^2   \equiv -\bar{m}^2~.
\label{w.11zz}
\end{equation}

Now, by normalization, we consider $\bar{m}^2 = 0$ for photons. Thus, for photons we obtain that 
\begin{eqnarray}
\label{Bardeen1}
&&\dot{t}= \frac{-2 a \mathcal{M}_b L+E r^3+a^2 E (2 \mathcal{M}_b+r)}{r ( r^2-2 \mathcal{M}_b r +a^2  )}~,\\
\label{Bardeen2}
&&\dot{\phi}= \frac{2 a \mathcal{M}_b E -2 \mathcal{M}_b L +r L}{r (r^2-2 \mathcal{M}_b r +a^2 )}~,\\
\label{Bardeen3}
&&\dot{r}^{2}= \frac{2 (L-a E)^2 \mathcal{M}_b+ (a^2 E^2-L^2)r+E^2 r^3}{r^3}~. 
\end{eqnarray}

The distance of the closest approach $r_0$ for the metric (\ref{metricBardeen}) can be obtained from $\dot{r}=0$, which yields 
\begin{equation}
\label{ipB}
\frac{r_0}{b}=\sqrt{1-\left(\frac{a}{b}\right)^2-\frac{2m}{r_0}\left(1-\frac{a}{b}\right)^2\left(\frac{r_0^2}{g_{\star}^2+r_0^2}\right)^{3/2}}~,
\end{equation}
where $b=L/E$ is the impact parameter. 

\begin{figure}[h]
\begin{center}
\includegraphics[width=0.4\textwidth]{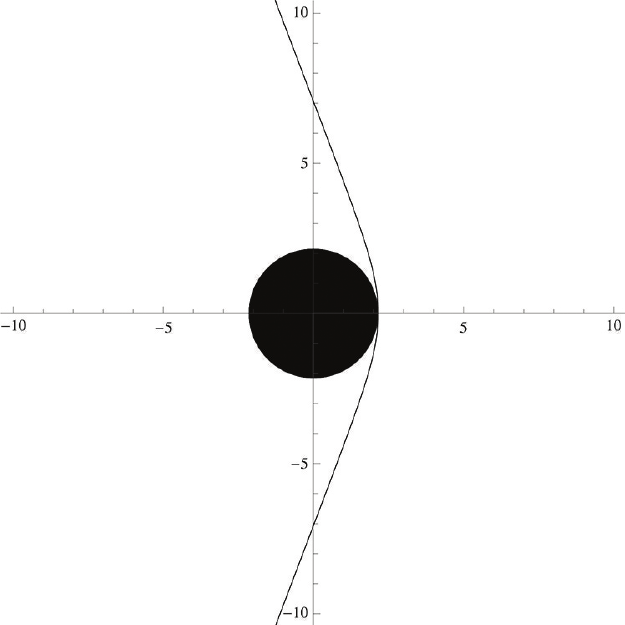}
\end{center}
\caption{The deflection of light in the background of a rotating regular Bardeen black hole with $E=2$, $L=5$, $m=0.3$, $g_{\star}=0.1$ and $a=0.1$. The circle corresponds to the closest approach ($r_0=2.154$) to the black hole.} \label{plotBardeen}
\end{figure}

Therefore, the bending angle Eq. (\ref{angle}) is given by
\begin{equation}
\hat{\alpha}=\frac{4m}{b}-\frac{8 g_{\star}^2 m}{b^3}\pm \frac{4ma}{b^2}+\mathcal{O}(a,m,g_{\star}^2 )
\end{equation}
where, similar to the previous case, we use the change of variables $u=r_0/r$, next, we substitute the impact parameter given by the Eq. (\ref{ipB}), then, we expand in Taylor series around $m$, $a$, and $g_{\star}$, and finally, we consider $r_0\approx b$. In Fig. \ref{plotBardeen} we plot the deflection of light in the background of a rotating Bardeen black hole by solving numerically the Eqs.(\ref{Bardeen2}) and (\ref{Bardeen3}). 

\section{Deflection angle by rotating regular Hayward
black hole}

In this section, we investigate the deflection angle by rotating Hayward regular black hole. The spacetime metric of the rotating Hayward regular black hole reads \cite{Ghosh}: 
\begin{equation}
ds^2=-\left(1-\frac{2\mathcal{M}_h r}{\Sigma}\right)dt^2-\frac{4 a\mathcal{M}_h r \sin^2\theta }{\Sigma }dt d\phi+\frac{\Sigma}{\Delta}dr^2+\Sigma d\theta^2+\left( r^2+a^2+\frac{2 a^2 \mathcal{M}_h r \sin^2\theta}{\Sigma}\right)\sin^2\theta d\phi^2~,
\end{equation} 
where
\begin{equation}
\Sigma=r^2+a^2\cos^2\theta~,
\end{equation}
\begin{equation}
\Delta=r^2-2\mathcal{M}_hr+a^2~,
\end{equation}
and in the equatorial plane the mass function is given by:
\begin{eqnarray}
\mathcal{M}_h &=& m\frac{r^3}{r^3+g^3}~, 
\end{eqnarray}
being $g$ the rotation parameter. We find the following expression for the optical metric:

\begin{equation}\label{77}
\mathcal{F}\left(r,\phi,\frac{\mathrm{d}r}{\mathrm{d}t},\frac{\mathrm{d}\phi}{\mathrm{d}t}\right)=\sqrt{\frac{r^4}{\Delta(\Delta-a^2)}(\frac{\mathrm{d}r}{\mathrm{d}t})^2+\frac{r^4 \Delta}{\Delta-a^2}(\frac{\mathrm{d}\phi}{\mathrm{d}t})^2}- \frac{2 m a r}{\Delta-a^2}\left( \frac{r^3}{r^3+g^3} \right)\frac{\mathrm{d}\phi}{\mathrm{d}t}~,
\end{equation}
with metric components:
\begin{eqnarray}\notag
\bar{g}_{rr}&=& - \frac{2 \left[-2 (m+\frac{r}{4}) \left(\sin^4\phi r^2 +b^2 \cos^2 \phi \right)^{3/2}+a m r^2 \sin^6 \phi    \right]}{r \left(\sin^4\phi r^2 +b^2 \cos^2 \phi \right)^{3/2}}+\frac{2m g^3 \left(a r^2 \sin^6\phi -2 \left(\sin^4\phi r^2 +b^2 \cos^2 \phi \right)^{3/2}\right)}{r^4 \left(\sin^4\phi r^2 +b^2 \cos^2 \phi \right)^{3/2}} ~,\\
 \bar{g}_{r\phi}&=& \frac{2 m a \cos^3\phi}{r\left(\frac{\sin^4\phi r^2 +b^2 \cos^2 \phi }{b^2}\right)^{3/2}}- \frac{2 m a \cos^3\phi g^3}{r^4\left(\frac{\sin^4\phi r^2 +b^2 \cos^2 \phi }{b^2}\right)^{3/2}}~,\\
 \bar{g}_{\phi \phi}&=&  \frac{2r \left[-2amr^2 \sin^6\phi-3a b^2 m \cos^2\phi \sin^2\phi +(m+\frac{r}{2}) \left(\sin^4\phi r^2 +b^2 \cos^2 \phi \right)^{3/2}  \right]}{\left(\sin^4\phi r^2 +b^2 \cos^2 \phi \right)^{3/2}}\\
 &-& \frac{2m g^3 \left[-2a r^2 \sin^6\phi -3 ab^2 \cos^2\phi \sin^2\phi +\left(\sin^4\phi r^2 +b^2 \cos^2 \phi \right)^{3/2}    \right]}{r^2 \left(\sin^4\phi r^2 +b^2 \cos^2 \phi \right)^{3/2}}~.
\end{eqnarray}

The determinant is given by
\begin{eqnarray}\notag
\det \bar{g}&=& r^2- \frac{6 r m \left[a \sin^2\phi-\sqrt{ \cos^4\phi r^2+(b^2-2r^2)\cos^2 \phi +r^2}    \right]}{\sqrt{ \cos^4\phi r^2+(b^2-2r^2)\cos^2 \phi +r^2}}\\
&+&\frac{6 g^3 m \left[a \sin^2\phi-\sqrt{ \cos^4\phi r^2+(b^2-2r^2)\cos^2 \phi +r^2}    \right]}{r^2\sqrt{ \cos^4\phi r^2+(b^2-2r^2)\cos^2 \phi +r^2}}~.
\end{eqnarray}

Our computation reveals the following relation
\begin{equation}
\mathcal{K}=-2\,{\frac {m}{{r}^{3}}}+20\,{\frac {m{g}^{3}}{{r}^{6}}}+\frac{36 a m \mathcal{H}(r,\phi)}{r^6}~,
\end{equation}
in which
\begin{eqnarray}\notag
\mathcal{H}(r,\phi)&=& \frac{\sin^2\phi}{\left(\sin^4\phi r^2 +b^2 \cos^2 \phi \right)^{7/2}}\Big[ \left( 5/4\,{g}^{3}{r}^{6}-1/6\,{r}^{9} \right)  \left( \sin \left( 
\phi \right)  \right) ^{12}\\\notag
&+&  \left( -1/12\,{g}^{3}{r}^{4}+1/12\,{r}^{7} \right) {b}^{2} \left( 
\sin \left( \phi \right)  \right) ^{10}+\left( \cos \left( \phi \right)  \right) ^{6}{b}^{6}{g}^{3}\\\notag
&+& 10/3\,{r}^{4}{b}^{2} \left( \cos \left( \phi \right)  \right) ^{2}
 \left( {g}^{3}+1/10\,{r}^{3} \right)  \left( \sin \left( \phi
 \right)  \right) ^{8}\\\notag
 &+& 7/6\,{r}^{3} \left( {g}^{3}-4/7\,{r}^{3} \right) {b}^{3} \left( \cos
 \left( \phi \right)  \right) ^{2} \left( \sin \left( \phi \right) 
 \right) ^{7}\\\notag
 &+& 1/3\,{r}^{2}{b}^{2} \left( \cos \left( \phi \right)  \right) ^{2}
 \left( {g}^{2}+gr+{r}^{2} \right)  \left( -5/2\,{r}^{2} \left( \cos
 \left( \phi \right)  \right) ^{2}+{b}^{2} \right)  \left( g-r
 \right)  \left( \sin \left( \phi \right)  \right) ^{6}\\\notag
 &+& 7/3\,{r}^{3} \left( {g}^{3}-4/7\,{r}^{3} \right) {b}^{3} \left( \cos
 \left( \phi \right)  \right) ^{4} \left( \sin \left( \phi \right) 
 \right) ^{5}\\\notag
 &+& 13/3\,{r}^{2} \left( {g}^{3}-{\frac {9\,{r}^{3}}{52}} \right) {b}^{4}
 \left( \cos \left( \phi \right)  \right) ^{4} \left( \sin \left( \phi
 \right)  \right) ^{4}\\\notag
 &+& 1/3\,r{b}^{5} \left( \cos \left( \phi \right)  \right) ^{4} \left( {g}
^{3}+1/2\,{r}^{3} \right)  \left( \sin \left( \phi \right)  \right) ^{
3}\\
&+& \left( 5/6\,{g}^{3}{r}^{2}-5/6\,{r}^{5} \right) {b}^{4} \left( \cos
 \left( \phi \right)  \right) ^{6} \left( \sin \left( \phi \right) 
 \right) ^{2}+2/3\,r{b}^{5} \left( \cos \left( \phi \right)  \right) ^
{6} \left( {g}^{3}+1/2\,{r}^{3} \right) \sin \left( \phi \right)
    \Big]~.
\end{eqnarray}

Going through the same procedure the deflection angle can be calculated by the following integral
\begin{equation}
\hat{\alpha} \simeq  -\int\limits_{0}^{\pi}\int\limits_{\frac{b}{\sin \phi}}^{\infty}\left( -2\,{\frac {m}{{r}^{3}}}+20\,{\frac {m{g}^{3}}{{r}^{6}}}+\frac{36 a m \mathcal{H}(r,\phi)}{r^6}\right)\,\sqrt{\det \bar{g}}\,\mathrm{d}r\,\mathrm{d}\phi ~.
\end{equation}

After we evaluate the first integral we find
\begin{eqnarray}\notag
\mathcal{I}_{1}&=&-\int\limits_{0}^{\pi}\int\limits_{\frac{b}{\sin \varphi}}^{\infty}\left( -2\,{\frac {m}{{r}^{3}}}+20\,{\frac {m{g}^{3}}{{r}^{6}}}\right)\sqrt{\det \bar{g}}\mathrm{d}r\mathrm{d}\phi\\
&=& \frac{4m}{b}-\frac{15 m \pi g^3}{8 b^4}~.
\end{eqnarray}

The second integral gives 
\begin{equation}
\mathcal{I}_2= -\int\limits_{0}^{\pi}\int\limits_{\frac{b}{\sin \phi}}^{\infty}\left( \frac{36 a m \mathcal{H}(r,\phi)}{r^6}\right)\,\sqrt{\det \bar{g}}\,\mathrm{d}r\,\mathrm{d}\phi=\pm \frac{4ma}{b^2}~.
\end{equation}

Consequently the total deflection angle of rotating Hayward regular black hole results: 
\begin{equation} 
\hat{\alpha} = \frac{4m}{b}-\frac{15 m \pi g^3}{8 b^4}\pm \frac{4ma}{b^2}~.
\end{equation}

\subsection{Geodesics equations}

The Lagrangian associated with the motion of particles in the equatorial plane ($\theta=\pi/2$ and $\dot{\theta}=0$) of a rotating regular Hayward
black hole is given by
%(\ref{tl1}) 
\begin{equation}
\label{LHayward}
  2\mathcal{L}= -\left(1-\frac{2\mathcal{M}_h }{r}\right)\dot{t}^2-\frac{4 a\mathcal{M}_h   }{r}\dot{t} \dot{\phi}+\frac{r^2}{\Delta}\dot{r}^2+\left( r^2+a^2+\frac{2 a^2 \mathcal{M}_h  }{r}\right) \dot{\phi}^2~,
\end{equation}
being $\dot{q}=dq/d\tau$, and $\tau$ is an affine parameter along the geodesic. Such as the previous analysis the Lagrangian (\ref{LHayward}) is
independent of the cyclic coordinates ($t,\phi$). So, their
conjugate momenta ($\Pi_t, \Pi_{\phi}$) are conserved. Then, the equations of motion are obtained from
$ \dot{\Pi}_{q} - \frac{\partial \mathcal{L}}{\partial q} = 0$, and yield
\begin{equation}
\dot{\Pi}_{t} =0 ,  \quad \dot{\Pi}_{\phi}=0~,
\label{w.11aaa}
\end{equation}
where $\Pi_{q} = \partial \mathcal{L}/\partial \dot{q}$
are the conjugate momenta to the coordinate $q$, and is given by
\begin{equation}
\Pi_{t} = -\left(1-\frac{2\mathcal{M}_h}{r}\right)\dot{t}-\frac{2a\mathcal{M}_h}{r}\dot{\phi} \equiv -E~, \quad \Pi_{r}=\frac{r^2}{\Delta}\dot{r}~, \quad \textrm{and}\quad \Pi_{\phi}
=-\frac{2a\mathcal{M}_h}{r}\dot{t}+\left(r^2+a^2+\frac{2a^2\mathcal{M}_h}{r} \right)\dot{\phi}\equiv L~,
\label{w.11ccc}
\end{equation}
where $E$ and $L$ are dimensionless integration constants  associated to each of them. Thus, the Hamiltonian is given by
\begin{equation}
\mathcal{H}=\Pi_{t} \dot{t} + \Pi_{\phi}\dot{\phi}+\Pi_{r}\dot{r}
-\mathcal{L}
\end{equation}
\begin{equation}
2\mathcal{H}=-E\, \dot{t} + L\,\dot{\phi}+\frac{r^2}{\Delta}\dot{r}^2\equiv -\bar{m}^2~.
\label{w.11zzz}
\end{equation}

Now, by normalization, we consider $\bar{m}^2 = 0$ for photons. So, we obtain   

\begin{eqnarray}
\label{Hayward2}
&&\dot{t}=\frac{r E (r^2+a^2)-2 a (L-aE) \mathcal{M}_h}{r (r^2-2 r \mathcal{M}_h+a^2)}~,\\
\label{Hayward1}
&&\dot{\phi}= \frac{2 a E \mathcal{M}_h+r L- 2 L \mathcal{M}_h}{r (r^2-2 r \mathcal{M}_h+a^2)}~,\\
\label{Hayward3}
&&\dot{r}^{2}=  \frac{r \left(E^2 \left(a^2+r^2\right)-L^2\right)+2 \mathcal{M}_h (L-a E)^2}{r^3}~. 
\end{eqnarray}

In this case, the distance of the closest approach $r_0$ for the metric (\ref{metricBardeen}) obtained from $\dot{r}=0$, yields 
\begin{equation}
\label{ipHar}
\frac{r_0}{b}=\sqrt{1-\left(\frac{a}{b}\right)^2-\frac{2m}{r_0}\left(1-\frac{a}{b}\right)^2\left(\frac{mr_0^3}{g^3+r_0^3}\right)}~,
\end{equation}
where $b=L/E$ is the impact parameter. 

\begin{figure}[h]
\begin{center}
\includegraphics[width=0.4\textwidth]{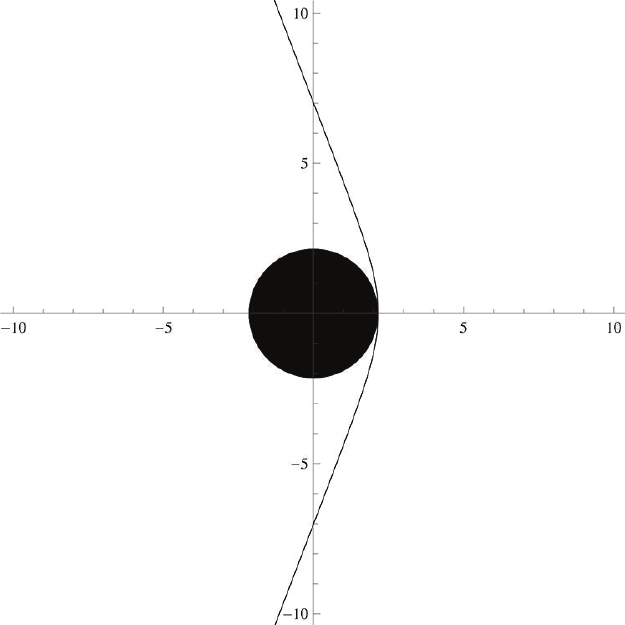}
\end{center}
\caption{The deflection of light in the background of a rotating Hayward black hole with $E=2$, $L=5$, $m=0.3$, $g=0.1$ and $a=0.1$. The circle corresponds to the closest approach ($r_0=2.153$) to the black hole.} \label{plotHayward}
\end{figure}

Therefore, the bending angle Eq. (\ref{angle}) yields
\begin{equation}
\hat{\alpha}=\frac{4m}{b}-\frac{15 m \pi g^3}{8 b^4}\pm \frac{4ma}{b^2}~.
\end{equation}
Here we use the change of variables $u=r_0/r$, we substitute the impact parameter given by the Eq. (\ref{ipHar}), we expand in Taylor series around $m$, $a$, and $g$; and finally, we consider $r_0\approx b$. In Fig. \ref{plotHayward} we plot the deflection of light in the background of a rotating Hayward black hole by solving numerically the Eqs. (\ref{Hayward1}) and (\ref{Hayward3}).  

\section{Conclusion}

In this paper, we have investigated the deflection angle of light by rotating regular black holes such as Ayon-Beato-Garc\'{i}a, 
%regular 
%black hole, 
%rotating regular 
Bardeen and  
%black hole and 
%rotating regular 
Hayward black hole. Starting from the physical metrics we have found the corresponding Rander-Finsler type metric which basically provides a way to compute the deflection angle in terms of GBT.  We have extended the Werner's geometric method by including the electric charge $Q$, magnetic charge $g_{\star}$, and deviation parameter $g$ which generalizes the expression for the Gaussian optical curvature, optical metric components, and finally the deflection angle.  

\begin{figure}[h]
\begin{center}
\includegraphics[width=0.45\textwidth]{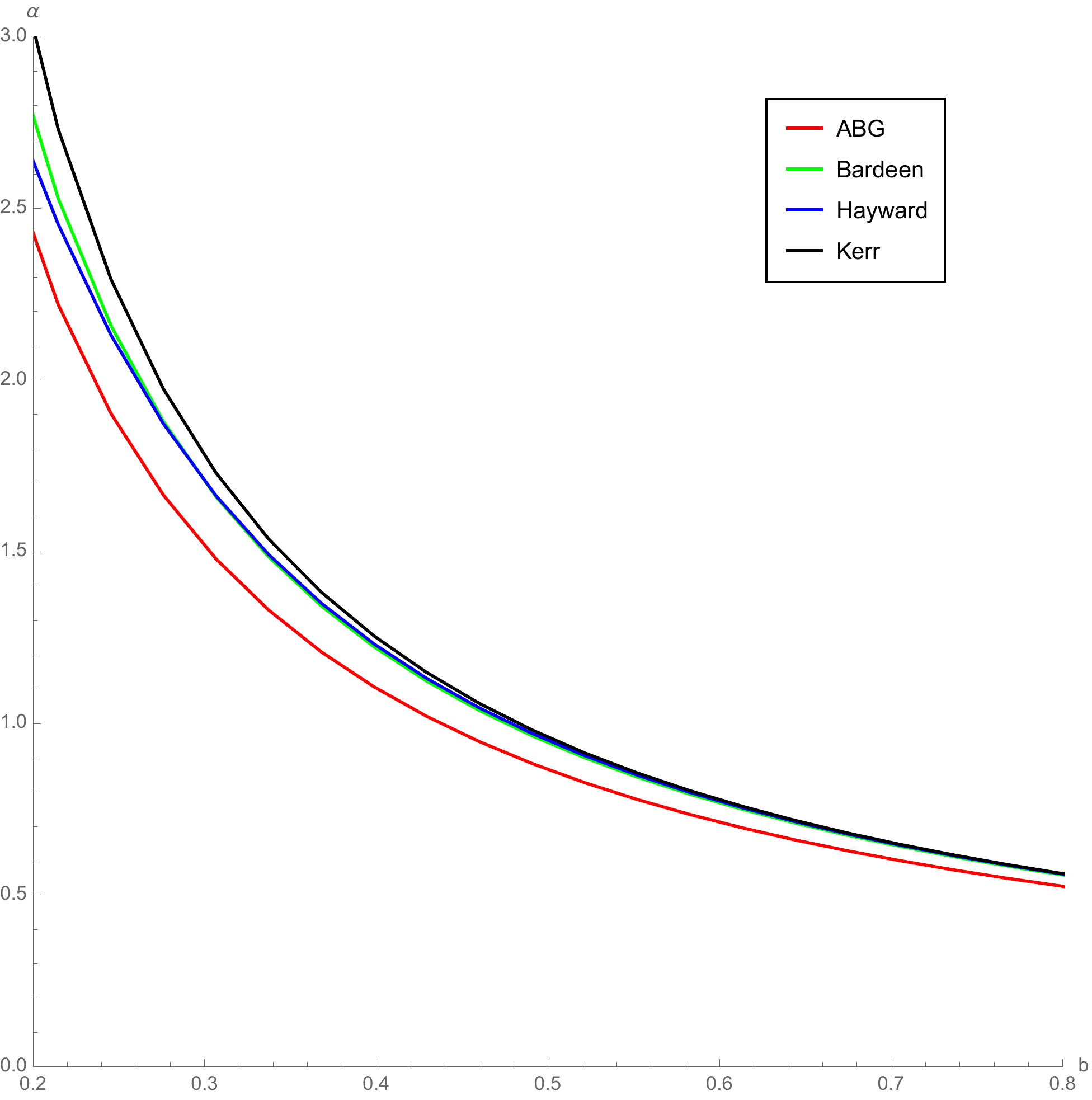}
\end{center}
\caption{The deflection of light by ABH, Bardeen, Hayward and Kerr black holes for the values of $m=Q=a=g=g_{\star}=0.1$.} \label{regular}
\end{figure}

In particular we have found that for the rotating  regular 
%ABG 
black holes the total deflection angles are %calculated as follows:
\begin{equation} 
\hat{\alpha}_{ABG} = \frac{4m}{b}-\frac{3 \pi Q^2}{4 b^2} \pm \frac{4ma}{b^2} +\mathcal{O}(Q^2,a,m)~,
\end{equation}
%for 
%In 
%the rotating regular Bardeen  black hole is
%case, the total deflection angle is obtained as follows:
\begin{equation}
\hat{\alpha}_{B}=
\frac{4m}{b}-\frac{8 g_{\star}^2 m}{b^3}\pm \frac{4ma}{b^2}+\mathcal{O}(a,m,g_{\star}^2 )~, 
\end{equation}
%and lastly for 
%, the total deflection angle of 
%the rotating regular Hayward  black hole is 
%found as follows: 
\begin{equation} 
\hat{\alpha}_{H} = \frac{4m}{b}-\frac{15 m \pi g^3}{8 b^4}\pm \frac{4ma}{b^2}+\mathcal{O}(a,m,g^3 ),
\end{equation}
for the ABG, Bardeen and Hayward regular black holes respectively.
%It is worth to mention that the total deflection angle of rotating Kerr black hole is \cite{werner,kerr}
Thus, as these black holes have in addition to the total mass and rotation parameter, different parameters as electric charge, magnetic charge, and deviation parameter the
%Newsworthy, we found that 
deflection of light has 
%the standard value of general relativity plus a 
correction terms coming from these parameters which generalizes the Kerr deflection angle
\begin{equation} 
\hat{\alpha}_{K} = \frac{4m}{b} \pm \frac{4ma}{b^2}.
\end{equation}
It is worth noting that, our results show that the deflection angle is smaller than the Kerr deflection angle, see Fig. \ref{regular}. In other words the contribution coming from the black hole parameters such as $Q$, $g_{\star}$ and $g$, is different in magnitude; however, in all three cases the light rays always bend outward the black holes which is indicated by the minus sign. In Fig. \ref{regular}, we show the behavior of the deflection angle of the light for the regular black hole geometries as function of the impact parameter. We observe that could exist discrepancies between predictions for the value of the deflection angle of the light for regular black holes for small values of the impact parameter, being the deflection angle smaller than the Kerr deflection angle. However, such discrepancy decreases when the impact parameter increases. In addition, we have checked our results of deflection angle using the geodesics formalism and we have shown to be exact in leading order terms.

It is important to realize that the agreement between these two methods breaks down for second order terms, for example, in the case of a rotating Bardeen regular black hole the geodesic approach gives  $\delta \hat{\alpha}_{B}=\mp 12 ma g_{\star}^2/b^4$,  whereas with the Gauss-Bonnet theorem one finds $\delta \hat{\alpha}_{B}= \mp 24 ma g_{\star}^2/b^4$. Such inconsistency is to be expected considering the fact that one must choose a different equation for the light ray $r_{\gamma}$ which incorporates the black holes parameters. We plan in the near future to extend our analytical analysis in the Gauss-Bonnet theorem to the second order terms to remove such an inconsistency.

\acknowledgments
This work is supported by Comisi\'on Nacional
de Ciencias y Tecnolog\'ia of Chile through FONDECYT Grant N$^{\textup{o}}$ 3170035 (A. \"{O}.), N$^{\textup{o}}$ 1170279 (J. S.) and by the Direcci\'{o}n de Investigaci\'{o}n y Desarrollo de la Universidad de La Serena (Y.V.). P. A. G. acknowledges the hospitality of the Universidad de La Serena where part of this work was undertaken.  A. \"{O}. is grateful to Prof. Douglas Singleton for hosting him at the California State University, Fresno and also thanks to Prof. Leonard Susskind and Stanford Institute for Theoretical Physics for hospitality.

\end{document}